\documentclass[conference]{IEEEtran}
\IEEEoverridecommandlockouts
\usepackage{cite}
\usepackage{amsmath,amssymb,amsfonts}
\usepackage{graphicx}
\usepackage{textcomp}
\usepackage{xcolor}
\def\BibTeX{{\rm B\kern-.05em{\sc i\kern-.025em b}\kern-.08em
    T\kern-.1667em\lower.7ex\hbox{E}\kern-.125emX}}

\usepackage{graphicx}
\usepackage{xcolor}
\usepackage{amssymb,amsmath,amsfonts}
\usepackage{tikz}
\usepackage{multirow}
\usepackage{booktabs}

\usepackage{blindtext}

\usepackage{multicol}
\usepackage{tikzscale}
\usepackage{numprint}
\npdecimalsign{.}
\nprounddigits{3}

\usepackage{parskip}

\usepackage{enumitem}

\usepackage{float}

\usepackage{pict2e}

\usepackage{siunitx}
\usepackage[english]{babel}

\newcommand{\Dd}{\mathbf D}

 \newcommand{\sd}{\mathbf{s}}                         
  
\newcommand{\XX}{\mathbf{x}}  
\newcommand{\EE}{\mathbf{e}}                       
\newcommand{\ZZ}{\mathbf{z}}                         
\newcommand{\YY}{\mathbf{y}} 
\newcommand{\UU}{\mathbf{u}} 
\newcommand{\BB}{\mathbf{b}}                       
\newcommand{\Ad}{\mathbf A}                         
 
\newcommand{\Fd}{\mathbf F} 
\newcommand{\Id}{\mathbf I}                         
\newcommand{\Cd}{\mathbf C}                         
\newcommand{\Ed}{\mathbf E}                         
\newcommand{\Hd}{\mathbf H}

  \newcommand{\Au}{\mathbf{A}_I}  

\newcommand{\herm}{{\scriptstyle \boldsymbol{\mathsf{H}}}}
\newcommand{\trans}{{\scriptstyle \boldsymbol{\mathsf{T}}}}

\usepackage[percent]{overpic}
\usepackage{algorithm,algcompatible,amsmath}

\algnewcommand\INPUT{\item[\textbf{Input:}]}%
\algnewcommand\PARAMETER{\item[\textbf{Parameters:}]}%
\algnewcommand\OUTPUT{\item[\textbf{Output:}]}%

\usepackage[
 colorlinks=false
]{hyperref}

\colorlet{lred}{red!80}
\colorlet{cmix}{blue!80!red} 
\colorlet{lgreen}{green!80}
\colorlet{lblue}{blue!80}

\numberwithin{theorem}{section}

 	\definecolor{amber(sae/ece)}{rgb}{1.0, 0.49, 0.0}
 	\definecolor{blue(ryb)}{rgb}{0.01, 0.28, 1.0}

    \usepackage{fancyhdr}

\fancypagestyle{specialfooter}{
  \fancyhf{}
  
  \fancyfoot[C]{\small{\textbf{$\copyright$ 2022 IEEE. Personal use of this material is permitted. Permission from IEEE must be obtained for all other uses, in any current or future media, including reprinting/republishing this material for advertising or promotional purposes, creating new collective works, for resale or redistribution to servers or lists, or reuse of any copyrighted component of this work in other works.}}}
}
 
\begin{document}

\title{Convolutional Dictionary Learning by
End-To-End Training of Iterative Neural Networks
}

\author{\IEEEauthorblockN{Andreas Kofler}
\IEEEauthorblockA{\textit{Physikalisch-Technische Bundesanstalt} \\
Braunschweig and Berlin, Germany \\
andreas.kofler@ptb.de}
\and
\IEEEauthorblockN{Christian Wald}
\IEEEauthorblockA{\textit{Department of Radiology} \\
\textit{Charité - Universitätsmedizin Berlin}\\
Berlin, Germany \\
christian.wald@charite.de}
\and
\IEEEauthorblockN{Tobias Schaeffter}
\IEEEauthorblockA{\textit{Physikalisch-Technische Bundesanstalt} \\
Braunschweig and Berlin, Germany \\
tobias.schaeffter@ptb.de}
\and
\IEEEauthorblockN{Markus Haltmeier}
\IEEEauthorblockA{\textit{Department of Mathematics} \\
\textit{University of Innsbruck}\\
Innsbruck, Austria \\
markus.haltmeier@uibk.ac.at} 
\and
\IEEEauthorblockN{Christoph Kolbitsch}
\IEEEauthorblockA{\textit{Physikalisch-Technische Bundesanstalt} \\
Braunschweig and Berlin, Germany \\
christoph.kolbitsch@ptb.de}
\and
}

\maketitle

\thispagestyle{specialfooter}

\begin{abstract}
Sparsity-based methods have a long history in the field of signal processing and have been successfully applied to various image reconstruction problems. The involved sparsifying transformations or dictionaries are typically either pre-trained using a model which reflects the assumed properties of the signals or adaptively learned during the reconstruction - yielding so-called blind Compressed Sensing approaches. However, by doing so, the transforms are never explicitly trained in conjunction with the physical model which generates the signals. In addition, properly choosing the involved regularization parameters remains a challenging task. Another recently emerged training-paradigm for regularization methods is to use iterative neural networks (INNs) - also known as unrolled networks - which contain the physical model. In this work, we construct an INN which can be used as a supervised and physics-informed online convolutional dictionary learning algorithm. We evaluated the proposed approach by applying it to a realistic large-scale dynamic MR reconstruction problem and compared it to several other recently published works. We show that the proposed INN improves over two conventional model-agnostic training methods and yields competitive results also compared to a deep INN. Further, it does not require to choose the regularization parameters and - in contrast to deep INNs - each network component is entirely interpretable.
\end{abstract}


\section{Introduction}
\label{sec:intro}
Recently, image reconstruction has become an important application area of machine learning (ML) and has attracted the interest of researchers in different communities. The most prominent type of ML-approaches are so-called iterative neural networks (INNs) or unrolled networks, see e.g.\ \cite{adler2017solving}, \cite{schlemper2017deep}, \cite{hammernik2018learning}. INNs correspond to iterative reconstruction schemes of finite length and typically consists of regularizing blocks, e.g.\ convolutional layers, and so-called data-consistency blocks which make use of the forward and adjoint operators. Integrating the physical model into the network reduces the maximum error-bound \cite{maier2019learning} and improves the network in terms of generalization properties \cite{kofler2021end}. In addition, INNs were reported to be the most robust with respect to adversarial attacks compared to model-agnostic networks \cite{Antun201907377}. Despite their success, \textit{deep} INNs are still black-boxes from a theoretical point of view. There is still no systematic way to design the employed convolutional blocks or to analyze them which raises concerns about their application in a sensitive field such as medical imaging. Nevertheless, the possibility to integrate the physical model into the learning process is a unique feature  of INNs which distinguishes them from other learning-based methods which, on the other hand,  often come with theoretical advantages such as convergence guarantees \cite{chun2017convolutional}.\\
Convolutional dictionary learning (CDL) is a well-established approach which has been extensively applied to  different tasks such as image denoising \cite{chun2017convolutional}, image inpainting \cite{papyan2017convolutional} and image reconstruction \cite{quan2016compressed}.
However, these methods typically learn the dictionary by solving the CDL-problem  and are not necessarily tailored to the reconstruction algorithm they are subsequently used with. In addition, properly tuning the regularization parameters is crucial (as we shall see later in the experiments) and is often a tedious and  difficult task.\\
In this work, we exploit the benefits of INNs - i.e.\ the inclusion of the physical model into the architecture - to construct a physics-informed and entirely interpretable ML-based  regularization method using CDL. The proposed method corresponds to an unrolled reconstruction scheme using a learned convolutional dictionary. The filters of the convolutional dictionary  as well as the involved regularization parameters - can then be trained in a supervised and physics-aware manner. We apply our method to a realistic large-scale dynamic MR image reconstruction problem  and demonstrate that i) the filters obtained by the NN-training yield better reconstructions compared to the ones obtained by decoupled pre-training of the filters as in \cite{chun2017convolutional} 
and \cite{liu2018first} and ii) that the proposed method shows competitive results compared to a state-of-the-art method  with deep iterative NNs \cite{schlemper2017deep}.

\section{Methods}
\label{sec:methods}
Our data measurement model corresponds to a general type of linear inverse problem
\begin{equation}\label{eq:inv_problem}
    \YY = \Ad \XX + \EE, 
\end{equation}
where $\Ad$ models the measurement process, $\XX$ denotes the  image, $\EE$ denotes Gaussian random noise and $\YY$ the measured data. The goal is to recover the unknown image $\XX$ from the observed measured data $\YY$. Because such problems can be  ill-posed for a variety of reasons, e.g.\ undersampling, poorly-conditioned systems or combinations thereof, the reconstruction requires the use of regularization methods which take additional information into account.

\subsection{Problem Formulation}
Similar to \cite{quan2016compressed}, we consider the following regularized reconstruction  problem  
\begin{equation}\label{eq:reco_problem}
\underset{\XX, \{\sd_k \}_k}{\min} \frac{1}{2}\|\Ad \XX - \YY\|_2^2 + \frac{\lambda}{2}   \Big\| \XX - \sum_{k=1}^K d_k \ast \sd_k \Big\|_2^2 +  \\ \alpha \sum_{k=1}^K \| \sd_k \|_1,
\end{equation}
where $\{d_k\}_{k=1}^K$ denotes a set of \textit{fixed} convolutional filters with $\| d_k \|_2=1$ for all $k$ and $\lambda, \alpha>0$.
Because the variables  $\sd_k$ are linked with the convolution operator within the $L_2$-norm and are also  present within the non-smooth $L_1$-norm, directly minimizing \eqref{eq:reco_problem} is difficult. We therefore  introduce auxiliary variables $\{\UU_k\}_k$ under the constraint $\UU_k = \sd_k$ for all $k$, yielding the equivalent problem
\begin{eqnarray}\label{xsu_reco_problem_constraint}
\underset{\XX, \{\sd_k \}_k, \{\UU_k \}_k}{\min} \frac{1}{2}\|\Ad \XX - \YY\|_2^2 + \frac{\lambda}{2}   \Big\| \XX - \sum_{k=1}^K d_k \ast \sd_k \Big\|_2^2  \\ \nonumber +  \alpha \sum_{k=1}^K \| \UU_k \|_1  \quad  \text{s.t.} \quad \forall k: \UU_k - \sd_k = \mathbf{0}.
\end{eqnarray}
Our aim is to set up a reconstruction algorithm based on \eqref{xsu_reco_problem_constraint} that defines a reconstruction network whose regularization parameters and filters $\{d_k\}_k$ are trained in an end-to-end manner.
\subsection{Reconstruction Algorithm}
The proposed solution strategy is to define an alternating minimization scheme by minimizing \eqref{xsu_reco_problem_constraint} with respect to one of the variables by keeping the others fixed. 
For fixed $\XX$,  problem \eqref{xsu_reco_problem_constraint} takes the form of 
\begin{eqnarray}\label{su_reco_problem_admm}
\underset{\{\sd_k \}_k, \{\UU_k \}_k}{\min}  \frac{\lambda}{2}   \Big\| \XX - \sum_{k=1}^K d_k \ast \sd_k \Big\|_2^2 +   \alpha \sum_{k=1}^K   \| \UU_k \|_1  \\ \nonumber \quad    \text{s.t.} \quad \forall k: \UU_k - \sd_k = \mathbf{0},
\end{eqnarray}
for which we use the alternating direction method of multipliiers (ADMM) \cite{boyd2011distributed} to update $\{\sd_k \}_k$ and $ \{\UU_k \}_k$. For $0\leq j\leq T$, the update-rules are given  by
\begin{eqnarray}\label{s_sub_problem}
\{\sd_k \}_k^{(j+1)} = \underset{\{\sd_k \}_k}{\arg \min} \,\frac{\lambda}{2}   \Big\| \XX - \sum_{k=1}^K d_k \ast \sd_k \Big\|_2^2 \\ \nonumber +   \frac{\beta}{2}  \sum_{k=1}^K\| \UU_k^{(j)} - \sd_k + \ZZ_k^{(j)}\|_2^2,
\end{eqnarray}
\begin{eqnarray}\label{u_sub_problem}
\{\UU_k \}_k^{(j+1)} = \underset{\{\UU_k \}_k}{\arg \min}\, \alpha \sum_{k=1}^K   \| \UU_k \|_1 \\  \nonumber +   \frac{\beta}{2}  \sum_{k=1}^K\| \UU_k - \sd_k^{(j+1)} + \ZZ_k^{(j)}\|_2^2,
\end{eqnarray}
\begin{eqnarray}\label{z_sub_problem}
  \{\ZZ_k \}_k^{(j+1)} = \{\ZZ_k \}_k^{(j)} + \{\UU_k \}_k^{(j+1)}- \{\sd_k \}_k^{(j+1)}  ,
\end{eqnarray}
where $\{\ZZ_k \}_k$ are the dual-variables of the ADMM algorithm.
\subsubsection{Update of $\{\sd_k\}_k$}
Updating $\{\sd_k\}_k$ corresponds to the most challenging problem and is achieved by solving the sub-problem in the Fourier-domain.
Define
\begin{equation}
    \XX^f:= \Fd \XX, \,
    d_k^f:= \Fd d_k,  \,
    \UU_k^f:= \Fd \UU_k, \,
    \sd_k^f:= \Fd \sd_k, \,
    \ZZ_k^f:= \Fd \ZZ_k.
\end{equation}
Then,  by making use of the Fourier-convolution theorem, updating $\{\sd_k\}_k$ according to \eqref{s_sub_problem} is equivalent to solving 
\begin{equation}\label{Fs_sub_problem}
\underset{\{\sd^f_k \}_k}{\min} \,   \frac{1}{2}   \Big\| \XX^f -  \sum_{k=1}^K  d_k^f \cdot  \sd_k^f \Big\|_2^2 +    \frac{\gamma}{2}  \sum_{k=1}^K\|   \UU_k^f -   \sd_k^f + \ZZ_k^f\|_2^2,
\end{equation}
where the multiplication $d_k^f \cdot  \sd_k^f$ is meant component-wise and $\gamma=:\beta/\lambda$.
Problem \eqref{Fs_sub_problem} can be re-written in a more compact form as 
\begin{equation}\label{Fs_sub_problem_op}
\underset{\sd^f }{\min} \,   \frac{1}{2}   \Big\| \XX^f -  \Dd^f \sd^f \Big\|_2^2 +    \frac{\gamma}{2}  \|   \UU^f + \ZZ^f -   \sd^f \|_2^2,
\end{equation}
where $\Dd^f:= [\Dd_1^f, \ldots, \Dd_K^f]=[\mathrm{diag}(d_1^f), \ldots, \mathrm{diag}(d_K^f)]$, $\sd^f:= [\sd_1^f, \ldots, \sd_K^f]^\trans, \, \UU^f:= [\UU_1^f, \ldots, \UU_K^f]^\trans$ and $\ZZ^f:= [\ZZ_1^f, \ldots, \ZZ_K^f]^\trans$. Solving problem \eqref{Fs_sub_problem_op} corresponds to solving a linear system 
\begin{equation}\label{lin_system_SM}
\big((\Dd^f)^\herm \Dd^f +\gamma\, \Id\big) \sd^f = (\Dd^f)^\herm \XX^f + \gamma\,(\UU^f + \ZZ^f)
\end{equation}
which, because of its special structure, has a computationally inexpensive closed-form solution which can be obtained using the Sherman-Morrison formula, see \cite{wohlberg2015efficient} for more details. 
Let $(\sd^f)^{\ast}$ denote the solution of \eqref{Fs_sub_problem_op}. Then we finally obtain $\sd_k^{(j+1)}:=\Fd^\herm (\sd_k^f)^\ast$ for all $k$ as the solution of \eqref{s_sub_problem}.
\subsubsection{Update of $\{\UU_k\}_k$ and $\{\ZZ_k\}_k$ }
Since the sub-problem in \eqref{u_sub_problem} is separable w.r.t.\ $k$, updating $\{\UU_k\}_k$ according to \eqref{u_sub_problem} is efficiently implemented by applying the soft-thresholding operator for all $k$, i.e.
\begin{equation}\label{eq:u_solution}
    \UU_k^{(j+1)}= \mathcal{S}_{\alpha/\beta} (\ZZ_k^{(j)} - \sd_k^{(j+1)}   ),
\end{equation}
where $\mathcal{S}_{\alpha/\beta}$ denotes the soft-thresholding operator with threshold $\alpha/\beta$. Updating the dual variables according to \eqref{z_sub_problem} is trivial.
\subsubsection{Update of $\XX$}
Last, for fixed $\{\sd_k \}_k$ and $\{\UU_k \}_k$, updating $\XX$ corresponds to solving the sub-problem 
\begin{equation}\label{x_sub_problem}
\underset{\XX}{\min} \frac{1}{2}\|\Ad \XX - \YY\|_2^2 + \frac{\lambda}{2}   \Big\| \XX - \sum_{k=1}^K d_k \ast \sd_k \Big\|_2^2 
\end{equation}
which can be achieved by solving the system $\Hd \XX = \BB$ with 
\begin{eqnarray}\
    \Hd =& \Ad^\herm \Ad + \lambda \Id \label{eq:x_system_rhs}\\
    \BB =& \Ad^\herm \YY + \lambda \sum_{k=1}^K d_k \ast \sd_k 
\end{eqnarray}
by any iterative solver, e.g.\ a conjugate gradient (CG)-method.
\subsection{Proposed Reconstruction Network} Note that the  updates of the respective variables describe a solution strategy for solving problem \eqref{xsu_reco_problem_constraint} where the set of convolutional filters $\{d_k\}_k$ as well as the parameters $\lambda$, $\alpha$ and $\beta$ are \textit{fixed}.  We construct a neural network whose blocks correspond to the operations required to solve the previously described sub-problems. Then, the filters $\{d_k\}_k$ as well as $\alpha, \beta $ and $\lambda$ are free parameters in the reconstruction network that can be learned by back-propagation on a set of input-target image pairs $\mathcal{D}=\{ (\XX_0^i, \XX_{\mathrm{f}}^i)_{i=1}^M\}$, where $\XX_{\mathrm{f}}$ denotes a ground-truth image which is used as label during training and $\XX_0:=\Ad^\sharp \YY$ denotes an initial guess of the image which is typically directly obtained from the measured data by some reconstruction operator $\Ad^\sharp$.\\
Typically, CDL algorithms impose a norm constraint on the filters, i.e.\ $\|d_k\|_2 = 1$,  to avoid the scaling ambiguity  between the filters and the sparse coefficient maps \cite{wohlberg2015efficient}, \cite{chun2017convolutional}. Since we train the filters by back-propagation, we project the filters onto the unit-sphere after each filter-update  by rescaling them by the corresponding $L_2$-norm.
\subsection{Implementation Details} Although the implementation of the single components of the proposed  network might seem straight-forward at a first glance, a closer look at the sub-problems reveals that some of the operations require careful treatment. First, the Fourier-convolution theorem used to obtain \eqref{Fs_sub_problem} only holds when circular padding conditions are assumed. Second, from \eqref{Fs_sub_problem}, one can see that the dimensionality of the filters in the Fourier-domain has to match the one of the image, which means that the filters have to be properly zero-padded. Third, to be able to use the Sherman-Morrison formula for solving \eqref{Fs_sub_problem_op}, one needs to construct the right hand side of the system \eqref{lin_system_SM}, where again, by proper circular and zero-padding, one needs to ensure that $(\Dd^f)^\herm$ and therefore implicitly $\Dd^\herm$ are indeeed the adjoint operators of $\Dd^f$ and $\Dd$, respectively. This means that they need to fulfill $\langle \Dd \mathbf{a}, \mathbf{b} \rangle = \langle  \mathbf{a}, \Dd^\herm \mathbf{b} \rangle$ and $\langle \Dd^f \mathbf{a}^f, \mathbf{b}^f \rangle = \langle  \mathbf{a}^f, (\Dd^f)^\herm \mathbf{b}^f \rangle$  for all $\mathbf{a}$ and $\mathbf{b}$.
Last, because we treat complex-valued images as two-channeled images and the filters are shared across the real and the imaginary part, these aspects also have to be taken into account when constructing all relevant operators. An implementation of our network is available at \texttt{\url{www.github.com/koflera/ConvSparsityNNs}}.
\section{Experiments}
\label{sec:experiments}
\begin{figure*}
\centering
\includegraphics[width=\linewidth]{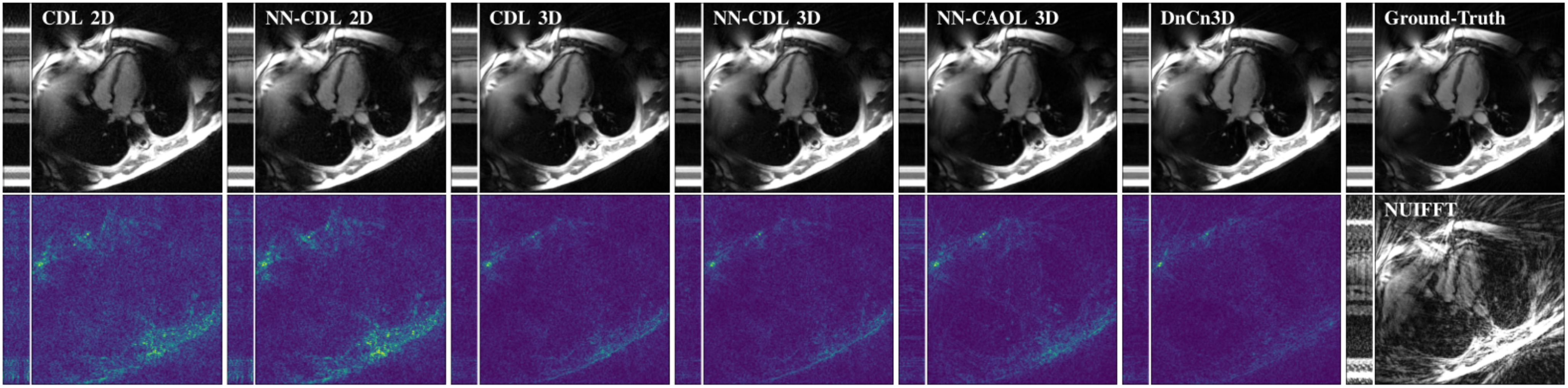}
\caption{An example of images reconstructed with the different reported methods: CDL 2D \cite{liu2018first}, NN-CDL 2D, CDL 3D \cite{chun2017convolutional},  NN-CDL 3D, NN-CAOL 3D \cite{kofler2022caol} and DnCn3D \cite{schlemper2017deep}. For CDL 3D, NN-CDL 3D and NN-CAOL 3D, we used $K=16$ filters with $k_f=7$,  and for NN-CDL 2D we used $K=96$ and $k_f=9$. Further, the target image and the initial reconstruction given by a non-uniform inverse FFT (NUIFFT) are shown.}\label{fig:reco_results}
\end{figure*}
We applied our proposed method to an accelerated  cardiac cine MR image reconstruction problem using radial  $k$-space trajectories. The operator $\Ad$ in  \eqref{eq:inv_problem} takes the form
\begin{equation}\label{eq:NUFFTOp}
\Ad:= (\Id_{N_c} \otimes \Ed) \Cd,
\end{equation}
for a complex valued image $\XX = [\XX_1, \ldots, \XX_{N_t}]^\trans \in \mathbb{C}^N$ with $N=N_x \times N_y \times N_t$, where $\Id_{N_c}$ denotes an identity operator and $\Cd$ contains the $N_c$ coil-sensitivity maps, i.e.\ $\Cd =  [\Cd_1,\ldots,\Cd_{N_c}]^\trans$, with $\Cd_j = \mathrm{diag}(\mathbf{c}_j,\mathbf{c}_j,\ldots,\mathbf{c}_j) \in \mathbb{C}^{N \times N}$ and $\mathbf{c}_j \in \mathbb{C}^{N_x \times N_y}$.  The operator $\Ed = \mathrm{diag}(\Ed_1, \ldots, \Ed_{N_t})$ consists of different 2D radial Fourier-encoding operators $\Ed_t$ which for each time point $t \in \{1,\ldots, N_t\}$ sample a 2D image $ \XX_t \in \mathbb{C}^{N_x \times N_y}$ along radial lines in Fourier-space. 
To accelerate the acquisition process, we only acquire a sub-portion of the $k$-space coefficients indexed by the set $I$ and denote the resulting operator by $\Au$. The $k$-space trajectories are chosen according to the golden-angle method and  $\Au$ was implemented using the library \texttt{TorchKBNufft} \cite{Muckley2020}. Further, we used $k$-space pre-conditioning, see \cite{kofler2022caol} for more details. 
\subsection{Dataset} We used a dataset of 15 healthy volunteers and four patients with a total of 216 cine MR images of shape $N_x\times N_y\times N_t=320 \times 320 \times 30$. The  data was split into 12/3/4 subjects (144/36/36 dynamic images) for training, validation and testing. The test set consisted of the four patients. As in \cite{kofler2022caol}, the initial $k$-space data with $N_c=12$ coils was retrospectively simulated using $N_{\theta}=36$ spokes per cardiac phase  and was corrupted by Gaussian noise with a standard deviation of $\sigma=0.02$ resulting in an approximate SNR 8.5. 
\subsection{3D vs 2D} 
Because training INNs can be computationally demanding, we also consider a variant of problem \eqref{eq:reco_problem}, where instead of using 3D filters, we use 2D filters and impose the regularization for each time-frame, i.e. the regularization term in \eqref{eq:reco_problem} changes to 
\begin{equation}\label{eq:reco_problem_2D}
\underset{ \{\sd_{k,t} \}_k}{\min}  \frac{\lambda}{2}   \sum_{t=1}^{N_t} \bigg( \Big\| \XX_t -  \sum_{k,t}^K d_k \ast \sd_{k,t} \Big\|_2^2   + \alpha \sum_{k=1}^K \| \sd_{k,t} \|_1 \bigg).
\end{equation}
This allows to use a larger number of filters during training at the price of not exploiting the temporal dimension for the regularization which is the one with the highest correlation among neighbouring pixels. Note that this formulation only changes the regularization term but not the reconstruction algorithm. Within the network,  the temporal dimension of the image $\XX$ can be dynamically merged back and forth to play the role of the batch-dimension depending on the stage of the reconstruction.
\subsection{Methods of Comparison  and Evaluation} Since our proposed approach is a method for training a convolutional dictionary, we compare it to two classical model-agnostic CDL-methods, which we refer to as  CDL 2D and CDL 3D. For this, we pre-trained the 2D and 3D convolutional filters using \cite{chun2017convolutional} and \cite{liu2018first} with $\alpha=1.0$ and $\alpha=0.2$, respectively. After having obtained the filters, we fixed them in our network and only trained  $\lambda$,  $\alpha$ and $\beta$ to exclude a possible change of performance which could be attributed to a sub-optimal choice of the regularization parameters. Further, we also compared with the method in  \cite{kofler2022caol}, which corresponds to the \textit{analysis operator} counterpart of this work and which we abbreviate by NN-CAOL 3D. We abbreviate our proposed methods by NN-CDL 3D and NN-CDL 2D, respectively. Last, we also compared our method to an adaptation of the deep cascade of CNNs presented in \cite{schlemper2017deep}, which we denote by DnCn3D, see \cite{kofler2022caol} for details. 
We trained all CDL-methods for different filter configurations $K=16,24$ with filters of shapes $k_f \times k_f \times k_f=7\times 7 \times 7$ for 3D and $K=64,96$ with $k_f\times k_f=9\times 9$ for 2D. Based on a hyper-parameter selection on the validation set, we finally used $K=16$ and $k_f=7$ for CDL 3D, NN-CDL 3D and CAOL 3D (see \cite{kofler2022caol}), while for CDL 2D and NN-CDL 2D, we used $K=96$ and $k_f=9$. The length of the networks and the number of CG-iterations for solving \eqref{eq:x_system_rhs} were set to $T=4$ and $n_{\mathrm{CG}}=12$ for all CDL-methods. Because problem \eqref{eq:reco_problem} is separable w.r.t.\ the temporal dimension, we trained all INNs on images only containing $N_t=4$ and $N_t=12$ cardiac phases for NN-CDL 2D and NN-CDL 3D, respectively.  The INNs were trained for 16 epochs by minimizing the squared $L_2$-error between the target image and the estimated one using ADAM with an initial learning rate of $5\cdot 10^{-4}$. 
All reconstructions were evaluated in terms of PSNR, NRMSE and SSIM which were calculated over a central squared ROI of $160 \times 160$ pixels.
\section{Results and Discussion}
\label{sec:results}
\begin{figure}[h!]
\centering
\includegraphics[width=0.75\linewidth]{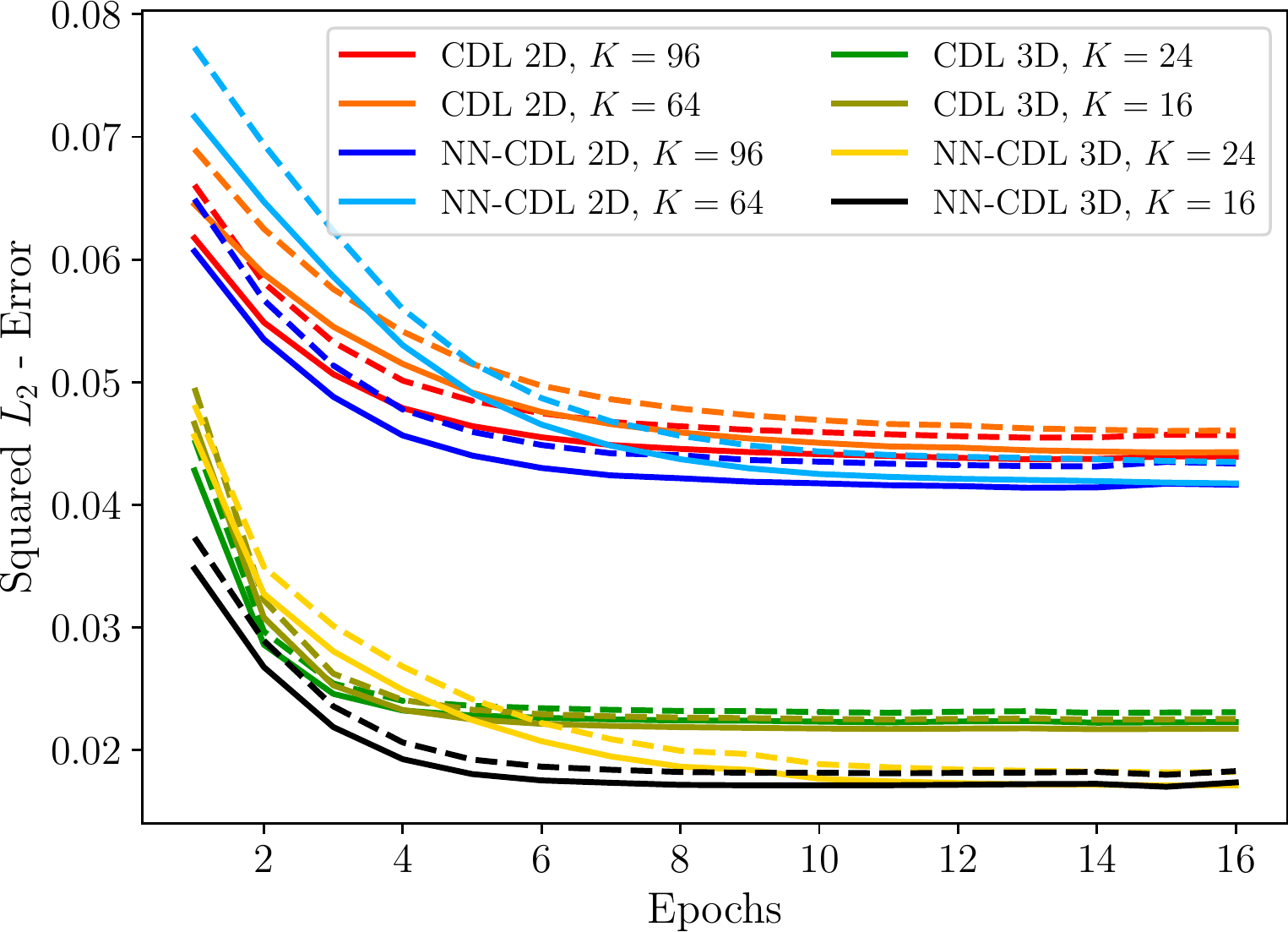}
\caption{Training- and validation-error (solid and dashed lines) for all CDL-methods. Training the convolutional filters with the proposed INN further reduces the validation-error compared to the de-coupled methods in  \cite{ liu2018first} and \cite{chun2017convolutional}. In addition, $\lambda$, $\alpha$ and $\beta$ can be trained as well. Note that for CDL 2D and CDL 3D, the filters were kept fixed and only $\lambda$, $\alpha$ and $\beta$ were trained. }\label{fig:training}
\end{figure}

Figure \ref{fig:reco_results} shows an example of images reconstructed by all  used methods. All methods yielded accurate reconstructions by successfully removing undersampling artefacts and noise. By comparing CDL 2D to NN-CDL 2D and CDL 3D to NN-CDL 3D, we see that using the proposed INN resulted in a more accurate reconstruction, by sligthly better removing residual noise. Further, by comparing NN-CAOL 3D with CDL 3D and NN-CDL 3D, we see that the CDL approach seems to give results which are marginally better than for the analysis operator approach. At no surprise, all sparsity-based methods are surpassed by DnCn3D which, however - being an INN with \textit{deep} convolutional blocks which also differ from iteration to iteration - contains eight to eleven times more trainable parameters than NN-CDL 2D and NN-CDL 3D, respectively, and is hardly  interpretable. Table \ref{tab:statistics} lists the average of the statistics obtained over all 2D images of each cardiac cycle of the test set. The table well-reflects the visually observed results. \\
Figure \ref{fig:training} shows the training- and the validation-error during the optimization of the proposed network. As we can see, training the filters by back-propagation further reduced the error compared to using the pre-trained filters obtained by  \cite{liu2018first} and \cite{chun2017convolutional}. The effect is clearly more pronounced for the 3D case. Further, we see how drastic the impact of not carefully chosen  $\beta$ and $\lambda$ can be on CDL 2D and CDL 3D. Note that, although CDL-2D and CDL-3D in the end achieve results which are comparable with the proposed method,  they only do so after a subsequent optimization of $\lambda$, $\beta$ and $\alpha$ using the proposed INN.  This demonstrates that the proposed INN can also be used as an algorithm for the selection of the regularization parameters.

\begin{table}[h!]
\centering
\caption{Quantitative results  for all different methods averaged over the test set.}
\begin{tabular}{@{}l|SSSc}
\hline
&  \textbf{NRMSE}  & \textbf{PSNR} &  \textbf{SSIM} & \textbf{$\sharp$ Params}\\
\hline
$\Au^\sharp$ & 0.36021885 & 36.449196 & 0.6336404088708495 & - \\
CDL 2D \cite{liu2018first}  & 0.121438004 & 45.5573 & 0.9422847189121768 & 3\,891 \\
NN-CDL 2D & 0.119256295 & 45.68295 & 0.9463142608393712 & 3\,891 \\
CDL 3D \cite{chun2017convolutional} & 0.080732286 & 49.226677 & 0.9724142688808683 & 2\,747 \\
NN-CDL 3D & 0.07887242 & 49.502735 & 0.9739793800511056 & 2\,747 \\
NN-CAOL 3D \cite{kofler2022caol} & 0.090448365 & 47.933865 & 0.9649090772386637 & 2\,746 \\
DnCn3D \cite{schlemper2017deep}& 0.07615194 & 50.191383 & 0.9727114839537564 & 31\,233 \\
\hline
\end{tabular}
\label{tab:statistics}
\end{table}

\begin{figure*}[h]
\centering
\includegraphics[width=\linewidth]{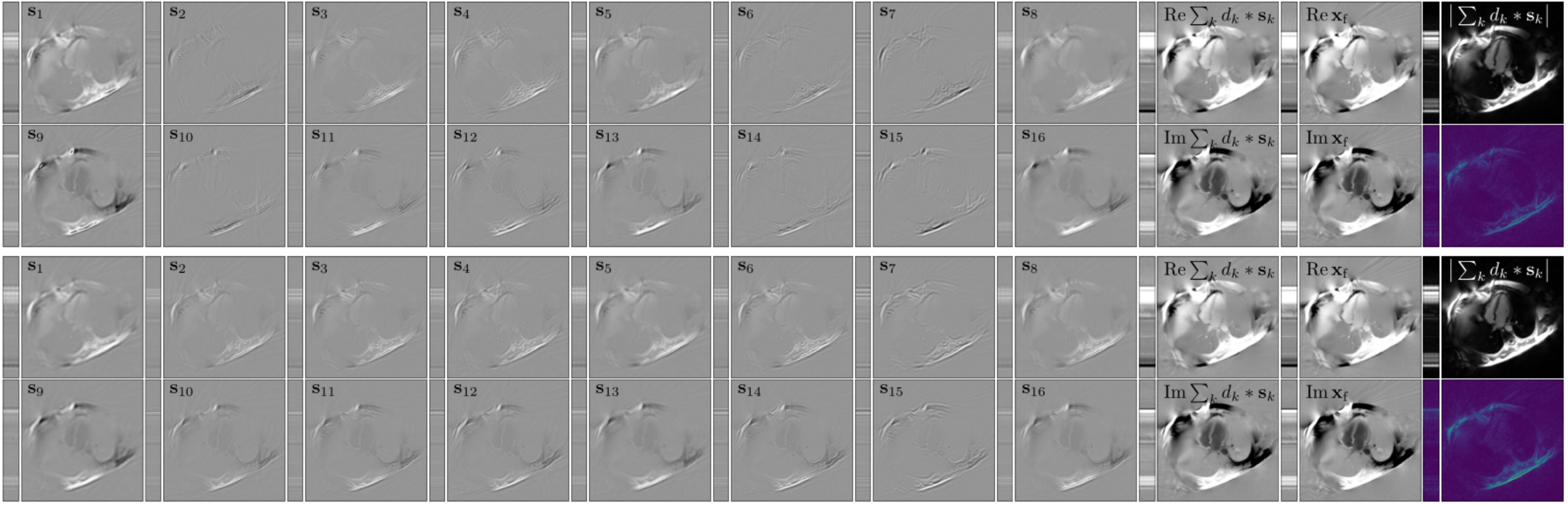}
\caption{Estimates of the sparse feature maps $\{s_k\}_k$ for $K=16$ and $k_f=7$ after having reconstructed the image solving problem \eqref{eq:reco_problem} using the 3D filters of NN-CDL 3D (top) and the ones of CDL 3D (bottom).  Further, the real and the imaginary parts of the approximation $\sum_k d_k \ast \sd_k$ as well as the magnitude image and the corresponding error-images are shown. The sparse feature maps are displayed in a window $[-15\,\sigma, 15\,\sigma$], where $\sigma$ denotes the standard deviation calculated over all 16  feature maps. As can be seen, the obtained sparse feature maps are relatively similar for both approaches. The filters obtained by the proposed NN-CDL slightly better approximate the ground-truth image and removed a larger portion of the undersampling artefacts.} \label{fig:sparse_approx}
\end{figure*}
Because the proposed method is interpretable, we can verify that the learned filters indeed serve the purpose they were designed for. Figure \ref{fig:sparse_approx} shows the estimates of $\{\sd_k\}_k$   at the last iteration of CDL 3D and NN-CDL 3D.  We can see that $\sum_k d_k \ast \sd_k$ indeed gives an approximation of the image and  the approximation using the filters of NN-CDL 3D exhibits a lower point-wise error compared to CDL 3D.
\section{Conclusion}
\label{sec:conclusion}
In this work, we have presented a physics-informed method for supervised training of a convolutional dictionary for image reconstruction applied to accelerated cardiac MRI. The proposed method consists of an iterative neural  network (INN) which is rigorously derived from a properly defined sparse reconstruction problem. The filters of the dictionary are trained in a supervised manner. We have compared our method to two unsupervised model-agnostic methods and showed competitive results where, most importantly, our method does not require the choice of the regularization parameters as they are jointly trained with the filters. The proposed INN also achieved  results comparable to its analysis operator counterpart and to a method based on deep INNs. On top, each learned component of the proposed reconstruction method is entirely interpretable in contrast to INNs with deep convolutional blocks.

\end{document}